# Optimized single-layer MoS$_2$ field-effect transistors by non-covalent functionalisation

HyunJeong Kim,‡[1,2] WungYeon Kim,‡[1,2] Maria O'Brien,[1] Niall McEvoy,[1] Chanyoung Yim,[3] Mario Marcia,[4] Frank Hauke,[4] Andreas Hirsch,[4] Gyu-Tae Kim,[2] and Georg S. Duesberg[1,3,a]

[1] CRANN&AMBER Centres and School of Chemistry, Trinity College Dublin, Dublin 2, Ireland

[2] School of Electrical Engineering, Korea University, Seoul 02841, South Korea

[3] Institute of Physics, EIT 2, Faculty of Electrical Engineering and Information Technology, Universität der Bundeswehr München, Neubiberg, Germany

[4] Chair of Organic Chemistry II, Friedrich-Alexander University of Erlangen-Nürnberg, Nikolaus Fiebiger Str. 10, 91058 Erlangen, Germany

E-mail: duesberg@unibw.de

‡These authors contributed equally to this work.

**Abstract**

Field-effect transistors (FETs) with non-covalently functionalised molybdenum disulfide (MoS$_2$) channels grown by chemical vapour deposition (CVD) on SiO$_2$ are reported. The dangling-bond-free surface of MoS$_2$ was functionalised with a perylene bisimide derivative to allow for the deposition of Al$_2$O$_3$ dielectric. This allowed the fabrication of top-gated, fully-encapsulated MoS$_2$ FETs. Furthermore, by the definition of vertical contacts on MoS$_2$, devices, in which the channel area was never exposed to polymers, were fabricated. The MoS$_2$ FETs showed high mobilities for transistors fabricated on SiO$_2$ with Al$_2$O$_3$ as top-gate dielectric. Thus, gate-stack engineering using innovative chemistry is a promising approach for the fabrication of reliable electronic devices based on 2D materials.

**Introduction**

Following the advent of graphene,[1] two-dimensional (2D) materials have been extensively examined as promising materials for nanoelectronics.[2–5] Unlike graphene, in which the absent bandgap limits its applications,[6–9] 2D semiconducting transition-metal dichalcogenides (TMDs), such as MoS$_2$, are considered as promising materials for future nanoelectronic devices.[10–14] To date, several methods have been introduced to obtain TMDs, each with benefits and drawbacks: Mechanical exfoliation delivers high-quality flakes[10,15] but suffers from scalability, while liquid-phase exfoliated materials have limited device performance.[16–19] Additionally, exfoliation leads to dispersed layer thicknesses, which give rise to varying properties.[15,20–22] Thermally assisted conversion (TAC) of pre-deposited metals or metal oxides yields a variety of large-scale TMD films with controllable thickness, however they are typically polycrystalline and defective.[23–25] Chemical vapour deposition (CVD) produces high-quality TMD crystals of predominantly monolayer thickness, however the growth of continuous, large-area films remains a challenge.[26–28] Given the associated advantages, CVD growth is the most suitable method for scalable fabrication of industry-relevant electronic devices.

Besides the challenges in synthesis, one of the major issues in fabrication of reliable devices with monolayer TMDs is control over the surface chemistry. Due to their monolayer nature the properties of 2D materials depend strongly on the environment. Adsorbates lead to doping via charge transfer, resulting in significant changes in the electrical properties. This effect is exploited in chemiresistors or ChemFETs for chemical sensing.[29,30] Furthermore, most monolayer TMDs are unstable and degrade under ambient conditions due to oxidation by water and/or oxygen.[31,32] This leads to deteriorated performance of FETs due to hysteresis and undefined doping. Therefore, it is important to passivate the 2D channels of devices for their stable operation. High-κ materials are considered to be excellent passivation layers and gate insulators. In general, they are suitable for low leakage and low power logic devices due to the high dielectric constant.[33] Furthermore, high-κ materials improve carrier mobility by reducing Coulomb scattering for nanostructures[34] or increasing the effective gate electric field.[35] $Al_2O_3$ is one of the most commonly used high-κ materials, often deposited using trimethylaluminum (TMA, $Al(CH_3)_3$) and an oxidation agent, water or ozone, by atomic layer deposition (ALD). However, the basal planes of 2D nanosheets, such as graphene or $MoS_2$, do not react with TMA due to the lack of dangling bonds or surface hydroxyl groups.[36–39] Thus, the formation of a seeding layer is required to perform ALD on the clean surface of 2D van der Waals crystals. We have shown that when perylene bisimides are deposited from the liquid phase onto graphene they form self-assembled monolayers (SAMs).[40,41] Specifically, -COOH and -OH units of the molecule react with TMA and promote the nucleation.[33,42,43] The same non-covalent functionalisation route was adopted for TMD layers,[44] wherein the perylene bisimide functional layer served as a stable seeding layer for $Al_2O_3$ deposition via ALD.

In this study we fabricated top-gated FETs with functionalised CVD-grown $MoS_2$. The perylene bisimide functional layer is used as a seeding layer for $Al_2O_3$ deposition, fabricating fully encapsulated $MoS_2$ FETs. We compared the devices fabricated by two different process flows. In both cases the electrical integrity of the TMD is maintained. Furthermore, the influence of polymer residue on device performance could be quantified.

**Methods**

**$MoS_2$ Growth**

$MoS_2$ samples were grown in a micro cavity in a two-zone CVD furnace at 700°C as described previously.[26] The samples were grown directly on $SiO_2$/Si (300 nm thick $SiO_2$, highly p-doped Si) substrates with $MoO_3$ and sulfur as solid precursors.

**Device Fabrication**

Back-gated $MoS_2$ FETs:

The source/drain electrodes on $MoS_2$ flakes were patterned by electron beam lithography (EBL), with electron beam resist (PMMA - A4, MicroChem) which was spin-coated onto the sample. EBL was performed using a Zeiss Supra 40 with a Raith EBL kit. After baking at 180 °C and developing with a MIBK:IPA (1:3) solution, the metal electrodes (Ti/Au, 5 nm/50 nm) were deposited by sputtering using a Gatan Precision Etching and Coating System (PECS).

Top-gated $MoS_2$ FETs Type 1:

Directly after the $MoS_2$ growth, perylene bisimide dissolved in aqueous buffer solution was deposited on the sample. Source/drain electrodes were defined by EBL as described above. The metal electrodes

(Ti/Au, 5 nm/50 nm) were deposited by using an electron beam evaporator (Temescal FC-2000). After lift-off, a 34 nm thick $Al_2O_3$ layer was then deposited on the $MoS_2$ channel region which was defined by EBL, using ALD (TP01, ATV Technology) with TMA and $H_2O$ precursors at 80°C. The $Al_2O_3$ thickness was measured by atomic force microscopy (AFM) as shown in Figure S1(a). Finally, the top-gate metal electrode (Ti/Au, 5 nm/50 nm), patterned by EBL, was sputtered (Gatan Precision Etching and Coating System).

Top-gated $MoS_2$ FETs Type 2:

Directly after the $MoS_2$ growth, perylene bisimide dissolved in aqueous buffer solution was deposited on the $MoS_2$. Subsequently, a 16 nm thick $Al_2O_3$ layer was deposited onto the samples by ALD at 80°C with TMA and $H_2O$ as precursors. The height profile of the first ALD-grown $Al_2O_3$ layer measured by AFM is shown in Figure S1(b). PMMA (A4, MicroChem) was then spin-coated on the $Al_2O_3$ layer and source/drain electrodes were defined by EBL. After development, the exposed $Al_2O_3$ layer was removed by 2.38% Tetramethylammonium hydroxide (TMAH) solution in $H_2O$ at 55°C, creating a vertical contact hole (VIA). Then metal electrodes (Ti/Au, 5 nm/50 nm) were deposited by electron beam evaporation, followed by lift-off. In order to prevent the possible leakage of the gate dielectrics during electrical measurements, an additional 24 nm of thick $Al_2O_3$ was deposited onto the gate region. The AFM height profile of the second ALD $Al_2O_3$ thickness is shown in Figure S1(c). In a subsequent step, the top gate electrode was defined by EBL and the gate metal electrode (Ti/Au, 5 nm/50 nm) was sputtered, followed by lift-off.

Raman spectroscopy was performed by using LabRam ARAMIS IR2 (HORIBA JOBIN YVON) and a WITec Alpha 300R using a 532 nm laser as excitation source. Thicknesses of $Al_2O_3$ and $MoS_2$ were measured using AFM (MFP-3D, Asylum Research). The topographic images of the $MoS_2$ surface were measured by AFM (Park Systems Park XE100). Scanning electron microscopy (SEM) imaging was done using a MIRA3 (TESCAN). Electrical measurements were conducted on a JANIS probe station connected to a Keithley 2612A source meter unit under vacuum (top-gated FETs: ~3.7 Torr, back-gated FETs: ~$2.3 \cdot 10^{-4}$ Torr) at room temperature. The substrate was connected to the ground during the electrical measurement of the top-gated FETs.

**Results**

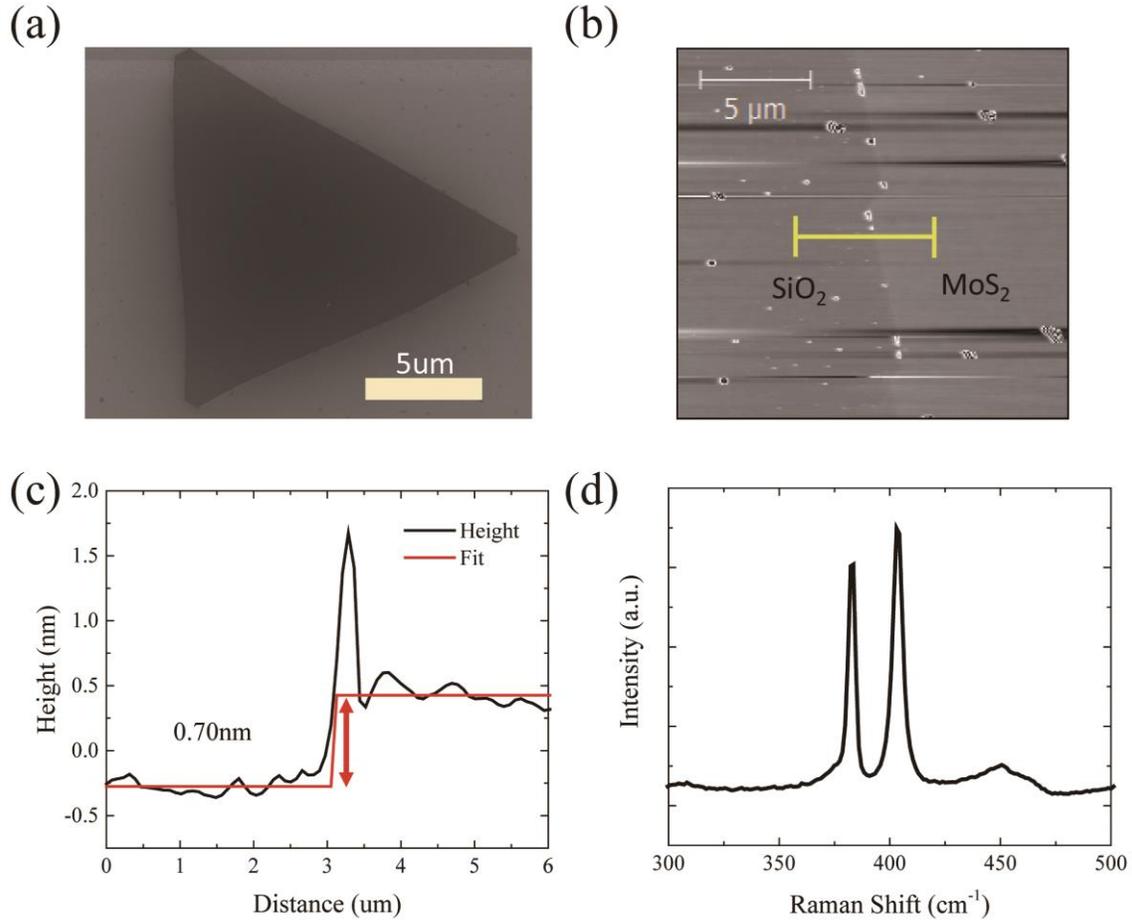

**Figure 1.** (a) SEM image of CVD-grown MoS$_2$. (b) AFM topography of CVD-grown MoS$_2$. (c) Average line profile along the marked line in (b). The step height between the flake and substrate is ~0.7 nm, confirming the monolayer nature of the triangular regions. (d) Raman spectrum of monolayer CVD-grown MoS$_2$.

The CVD growth yields randomly distributed, monolayer MoS$_2$ regions, e.g. flakes, which are mostly triangular in shape and extend over several micrometers. In Figure 1 typical results of CVD-grown MoS$_2$ on SiO$_2$ are presented. In Figure 1(a) a SEM image of the triangular shaped MoS$_2$ is shown. AFM measurements confirmed the monolayer nature of the triangular regions, as shown in Figure 1(b) and (c). A Raman spectrum of an as-grown MoS$_2$ flake is shown in Figure 1(d), the in-plane (E') and out-of-plane (A'$_1$) peaks occur at 383 and 403 cm$^{-1}$, respectively, which is consistent with monolayered MoS$_2$. Some bilayer and/or multilayer formation also can take place in the seeding regions of the flakes as shown in the inset of Figure 2(a) (line region of optical image).

To reveal the electrical properties of the CVD-grown MoS$_2$, flakes were contacted by EBL-defined electrodes as shown in figure S2(a). Using the substrate as the back gate, a FET with an MoS$_2$ channel was defined as schematically shown in the inset of figure S2(a). The FET shows on/off ratios with an order of 10$^3$, and the average field-effect mobility of the device is 0.66 cm$^2$/Vs for forward sweep and 0.79 cm$^2$/Vs for reverse sweep. These relatively low mobility values can be attributed to the scattering of carriers at the surface and the SiO$_2$ substrate.[45] In order to improve the performance, FETs in which the

MoS$_2$ channel is encapsulated and the channel has a top-gate electrode, separated by a high-κ oxide for effective modulation, can be fabricated. The gate dielectric deposition is ideally realised in a non-destructive and scalable manner by ALD. However, generally ALD on clean 2D van der Waals materials is challenging, due to the absence of seeding sites such hydroxyl or carbonyl oxides.[39,46–48]

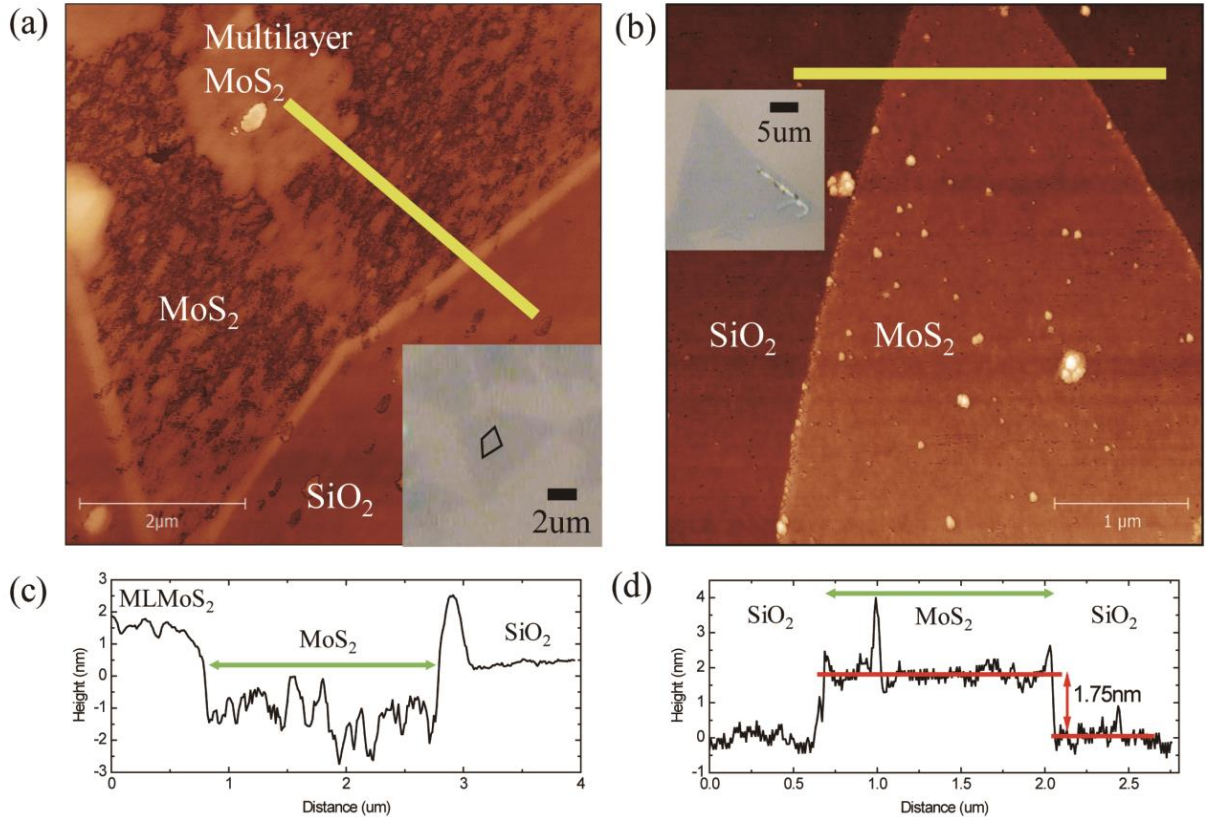

**Figure 2**. (a) AFM topography of pristine MoS$_2$ after Al$_2$O$_3$ deposition by ALD. A very rough surface due to Al$_2$O$_3$ islands can be seen on the MoS$_2$ regions. Inset: Optical image of the investigated region (b) AFM topography of MoS$_2$ functionalized by perylene bisimide after Al$_2$O$_3$ deposition by ALD. Inset: Optical image of the investigated region (c) Line profile along the marked line in (a). The MoS$_2$ lies in average lower than that in the SiO$_2$ substrate. (d) Line profile along the marked line in (b). The MoS$_2$ lies higher than that in the SiO$_2$ substrate.

AFM measurements were carried out to investigate the Al$_2$O$_3$ deposition on MoS$_2$. As also shown in figure 2 the triangular MoS$_2$ flakes consist of monolayers. There is some double and multilayer formation at the seeding region at the centre of the grain as shown in the inset of Figure 2(a) (line region). Figure 2(a) shows an AFM topography image of an ALD-Al$_2$O$_3$ layer deposited on a MoS$_2$ flake with 45 cycles of TMA/H$_2$O at 80 °C. Figure 2(c) shows the line profile of the yellow line of Figure 2(a). The step height between monolayer and substrate is approximately 0.34 nm. The MoS$_2$ lies on average lower than the SiO$_2$ substrate, and the monolayer surface is very rough. This is attributed to inhomogeneous and imperfect deposition of Al$_2$O$_3$ on the monolayer. Evidently, on the monolayer MoS$_2$ some Al$_2$O$_3$ island growth has taken place, but no continuous, complete coverage is reached. The step height between multilayer and substrate is approximately 1.3 nm, and this is close to the thickness of bilayer MoS$_2$. This indicates that in the multilayer region Al$_2$O$_3$ growth took place. Also, Al$_2$O$_3$ deposition can be seen at the edges of the monolayer MoS$_2$ flakes resulting in high step at the edge of the MoS$_2$ flake. This is attributed to a higher abundance of reactive site such as dangling bonds and defects at the edges of MoS$_2$ layers acting is anchor sites for the deposition of Al$_2$O$_3$.[44] Figure 2(b) shows an AFM topography image of an ALD-Al$_2$O$_3$ layer on a perylene bisimide-covered MoS$_2$ flake with 45 cycles of TMA/H$_2$O at 80 °C. In contrast to Figure 2(a), the Al$_2$O$_3$-MoS$_2$ surface in Figure 2(b) is uniform and lies higher than the SiO$_2$ substrate. Thus, the perylene bisimide layer acts as a seed for ALD growth, resulting in a homogenous and continuous Al$_2$O$_3$ layer. Figure 2(d) shows the line profiles of Figure 2(b) from the MoS$_2$ flake to the substrate. The step height between monolayer and substrate is approximately 1.75 nm, and this exceeds the monolayer thickness. The additional thickness is attributed to the perylene bisimide layer. Thus, the perylene bisimide functionalisation allows the ALD of dielectrics on clean dangling-bond-free TMD surfaces.

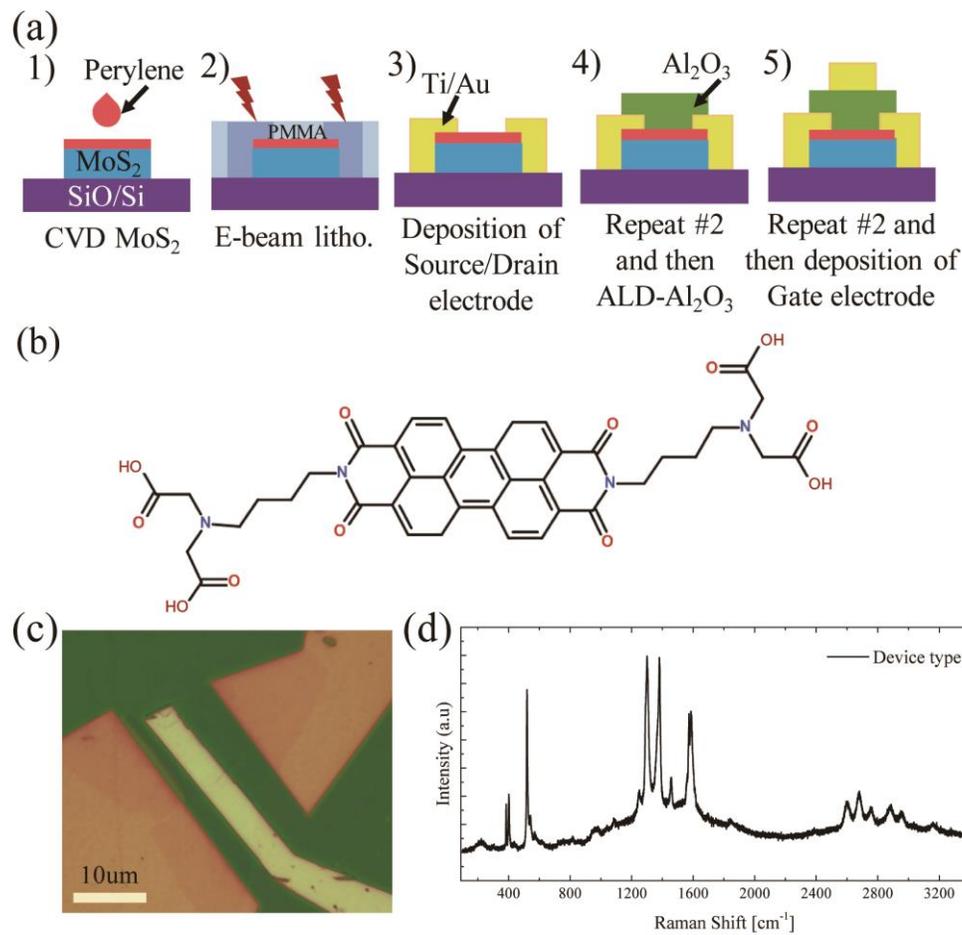

**Figure 3**. Top-gated MoS$_2$ FET type 1: (a) Schematic of process flow. (b) Chemical structure of the perylene bisimide derivative. (c) Top view optical image of fully fabricated device. (d) Raman spectrum of the channel area of a fully fabricated device.

Having optimized the dielectric deposition, we investigated the viability of our non-covalent functionalisation to optimize the gate-stack formation of 2D material FETs. The process flow to yield fully encapsulated top-gated MoS$_2$ FETs (device type 1) is shown in the schematic Figure 3(a). It follows the standard device fabrication processes with E-Beam resist (PMMA) deposition directly onto the whole substrate; however in our case the PMMA was deposited after deposition of the perylene bisimide (Figure 3(b)). Source/drain electrodes were patterned by EBL and deposited by evaporation. After lift-off, the $Al_2O_3$ gate dielectric was deposited on the entire $MoS_2$ channel region defined by EBL. This was followed by the deposition of the top gate electrode, after another lithographic step. An optical image of one device is shown in Figure 3(c). Raman spectroscopy was used to confirm the presence of $MoS_2$ monolayer and perylene SAM after completion of the gate stack. Figure 3(d) shows the peaks at positions at ~384 cm$^{-1}$ and 401 cm$^{-1}$ for the completed device, corresponding to the E' and A'$_1$ modes of $MoS_2$, respectively. Furthermore, the spectra exhibit several peaks in the region of 1300 cm$^{-1}$ to 1600 cm$^{-1}$ typical for perylene bisimide. This underlines the remarkable stability of the perylene SAM, it withstands the polymer removal with acetone and the ALD process.

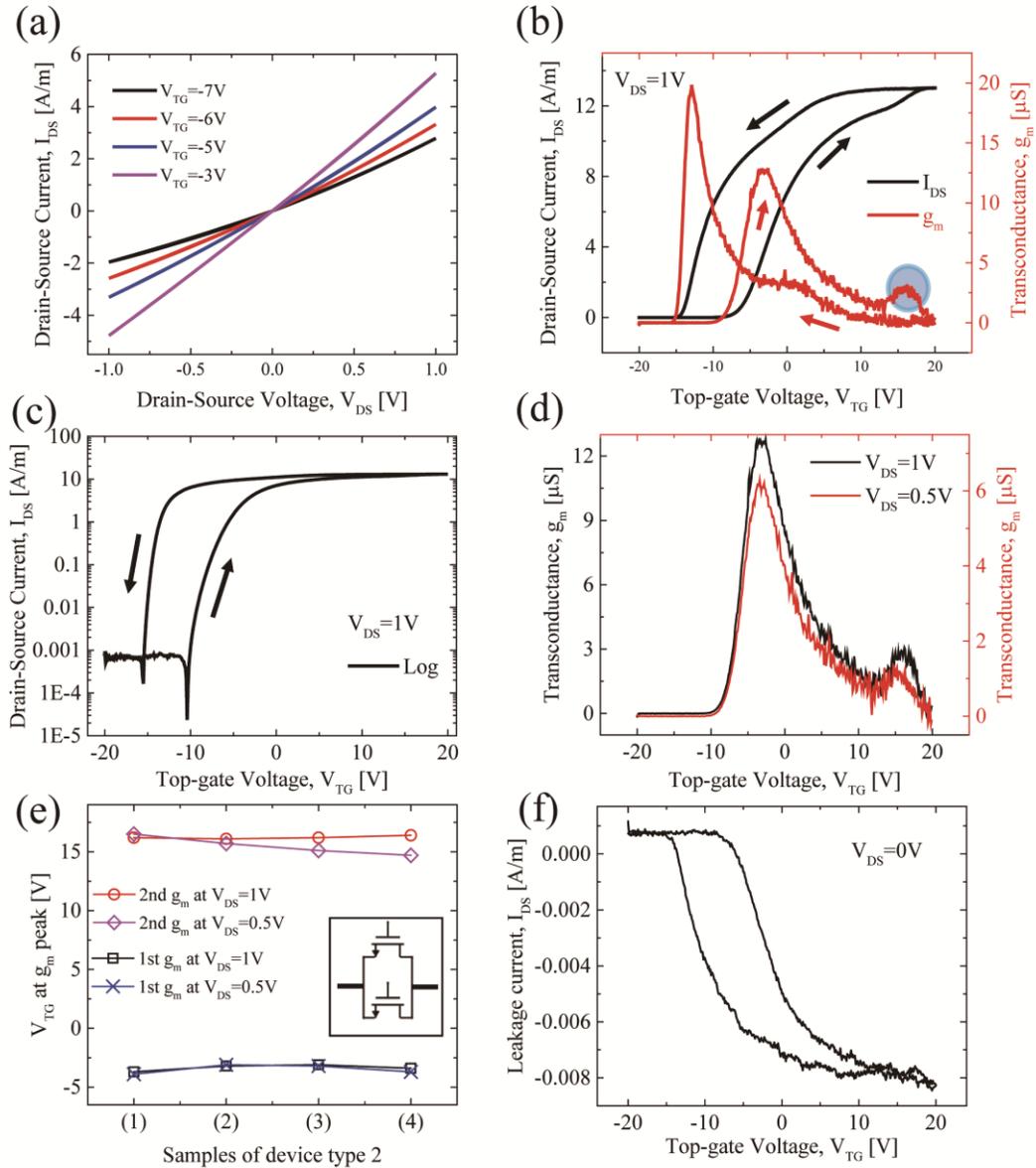

**Figure 4.** Electrical characterisation of device type 1: (a) Output characteristics. (b) Transfer characteristics (black line) and transconductance (red line) curves for various top-gate voltages at $V_{DS}$=1 V. The blue circle highlights the second transconductance peak. (c) Logarithmic plot of transfer

characteristics. (d) Transconductance curves at $V_{DS}=0.5V$ (red line) and 1V (black line). (e) Top-gate voltages for the first and second transconductance peaks of type 1 devices at $V_{DS}=0.5V$ and 1V. Inset: equivalent circuit model with two transistors connected in parallel. (f) Plot of leakage current versus top-gate voltage.

Figure 4(a) shows the $I_{DS}$-$V_{DS}$ output characteristics at various top-gate voltages ($V_{TG}$) of device type 1. The $SiO_2$ substrate of the device was grounded during all measurements, reducing possible capacitive coupling between top- and back-side dielectrics.[49] Linear behavior is observed at each $V_{TG}$, which indicates that the integration route yields good contacts and that the perylene bisimide layer does not significantly affect contact properties between $MoS_2$ and the electrodes. Figure 4(b) shows the typical $I_{DS}$-$V_{GS}$ transfer characteristics of device type 1. This exhibits a counter clockwise hysteresis. The reverse sweep of the top-gate voltage, $V_{TG}$, exhibits a higher conductivity compared with the forward sweep of $V_{TG}$. The counter clockwise hysteresis can be attributed to the positive mobile charges in $Al_2O_3$.[50,51] In the forward $V_{TG}$ sweep, positive mobile charges in $Al_2O_3$ move to the $Al_2O_3$/(perylene bisimide)/$MoS_2$ interface. In the reverse $V_{TG}$ sweep, the positive mobile charges located near the $MoS_2$ surface induce an additional electrical field, and this leads to a lower threshold voltage ($V_{TH}$) and increases current. Additionally, we compared transfer characteristics between two back-gated $MoS_2$ FETs, with and without a perylene bisimide layer on the $MoS_2$ channels. As shown in figure S2(c) and (f) of the Supporting Information, both devices show a similar hysteresis trend, indicating that the perylene bisimide layer does not play an important role in the observed hysteresis. Sub-threshold swing is 283 mV/decade for the forward sweep and 214 mV/decade for the reverse sweep. The field-effect mobility ($\mu_{FE}$) is calculated by the transfer characteristic using the following equation:

$$\mu_{FE} = \frac{Lg_m}{WC_{Al_2O_3}V_{DS}} \tag{1}$$

where $C_{Al2O3}$ denotes gate dielectric capacitance, $g_m$ denotes transconductance, $V_{DS}$ denotes drain-source voltage, L denotes channel length, and W denotes channel width. The channel shape does not often correspond to a rectangle, and thus the channel width is obtained by dividing the total channel area by the channel length. The $\mu_{FE}$ of the device is 21.4 cm$^2$/Vs for the forward sweep and 33.4 cm$^2$/Vs for the reverse sweep. The red line in Figure 4(b) indicates the $g_m$ of the device. The slope of drain/source current significantly increases until the $g_m$ reaches a peak, and then decreases when $V_{TG}$ increases. However, the device exhibits an intriguing second $g_m$ peak for the forward $V_{TG}$ sweep (blue-circled region) which is consistent with a small hump in the transfer characteristics at the same $V_{TG}$. Such second $g_m$ peak was also observed at various $V_{DS}$ (Figure 4(d)) in four out of five samples. Interestingly, as shown in Figure 4(e), the second $g_m$ peak appears at the similar $V_{TG}$ (~16 V) in all four samples regardless of $V_{DS}$, indicating that there is a common reason for the second $g_m$ peak with reproducibility. Similar signatures in the transfer characteristics were observed in SOI MOSFETs,[52–56] polysilicon thin film transistors,[57] and gate injection GaN-based transistors.[58] Even in the case of previous studies of SOI MOSFETs, which are more optimized than the $MoS_2$ FETs studied here, the origin of additional transport carriers varied depending on the device structure and materials. Thus, the origin of second $g_m$ peak cannot be exactly determined at this stage. However, as shown in Figure 4(b), a slight increase in current was observed at $V_{TG}$ of the second $g_m$ peak. As discussed in previous studies,[52–58] the injection of additional transport carriers could be considered as one of the reasons for the second $g_m$ peak. In the case of device type 1 the leakage current (see Figure 4(f)), obtained by measuring the $I_{DS}$ at $V_{DS} =0$ V under a $V_{TG}$ sweep, can be ruled out as a source of the additional carriers, since it is too low to affect the transfer characteristics. The devices can be expressed using an equivalent circuit model, composed of a main transistor connected to a parasitic transistor in parallel, as shown in the inset of Figure 4(e). In the equivalent circuit model, the threshold voltage of the main transistor differs from that of the parasitic transistor. The second $g_m$ peak value is small compared to the first $g_m$ peak, as shown in Figure 4(b). This is because the current generated after the parasitic transistor turns on is low. Thus, the second $g_m$ peak can be preliminarily attributed to the presence of polymer residue that acts as a parasitic transistor.

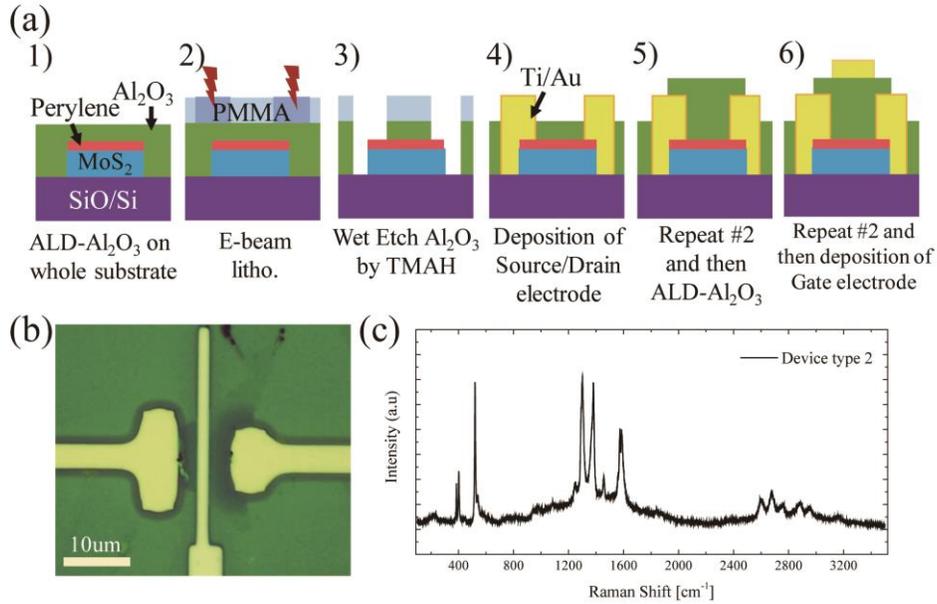

**Figure 5.** Top-gated MoS$_2$ FET type 2: (a) Schematic of process flow. (b) Top view optical image of fully fabricated device. (c) Raman spectrum of the channel area of a fully fabricated device.

A second more advanced process flow avoiding any contact of the channel region with polymer resist was developed. As shown in Figure 5(a), for device type 2 we deposited the Al$_2$O$_3$ layer on the whole substrate directly after perylene bisimide functionalisation. The source/drain electrodes were patterned by EBL, effectively creating contact holes by wet etch of the Al$_2$O$_3$ layer with an etch stop on the MoS$_2$. This realisation of vertical interconnects (VIAs), with metal evaporation to contact 2D materials is an important step to their successful integration. Importantly, this process flow has an advantage that the MoS$_2$ channel is never in contact with polymer resist. The presence of perylene between the Al$_2$O$_3$ and the MoS$_2$ was confirmed by using Raman spectroscopy. The spectra were taken after device fabrication was completed. Like device type 1, the typical signatures for MoS$_2$ at ~385 cm$^{-1}$ and 405 cm$^{-1}$ and perylene at 1300 cm$^{-1}$ to 1600 cm$^{-1}$ are observed as shown in Figure 5(c).

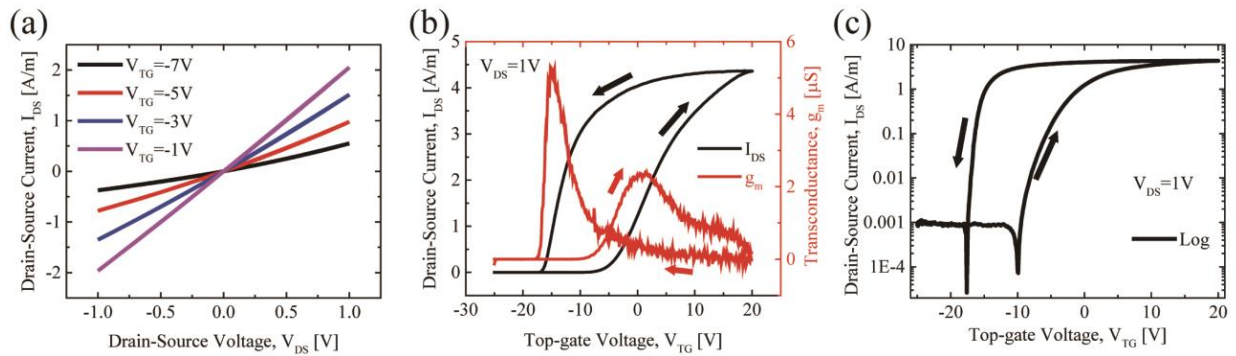

**Figure 6**. Electrical characterisation of device type 2**:** (a) Output characteristics. (b) Transfer characteristics (black line) and transconductance (red line) for various top gate voltages at $V_{DS}$=1 V. (c) Logarithmic plot of transfer characteristics.

In Figure 6(a), the output characteristics of device type 2 are shown. They exhibit a linear behavior like device type 1, indicating that the contacts between the monolayer MoS$_2$ and electrodes, which were defined by VIA etching and filling, were well established. This is an important achievement for the integration of 2D materials. Figure 6(b) shows transfer characteristics and transconductance of the device. A counter clockwise hysteresis appears, similar to the device type 1. Sub-threshold swing is 255 mV/decade for the forward sweep and 224 mV/decade for the reverse sweep. The $\mu_{FE}$ of device was extracted to be 22 cm$^2$/Vs for the forward sweep and 48.7 cm$^2$/Vs for the reverse sweep. Unlike device type 1, the second $g_m$ peak was not observed in device type 2, as shown in red line of Figure 6(b).

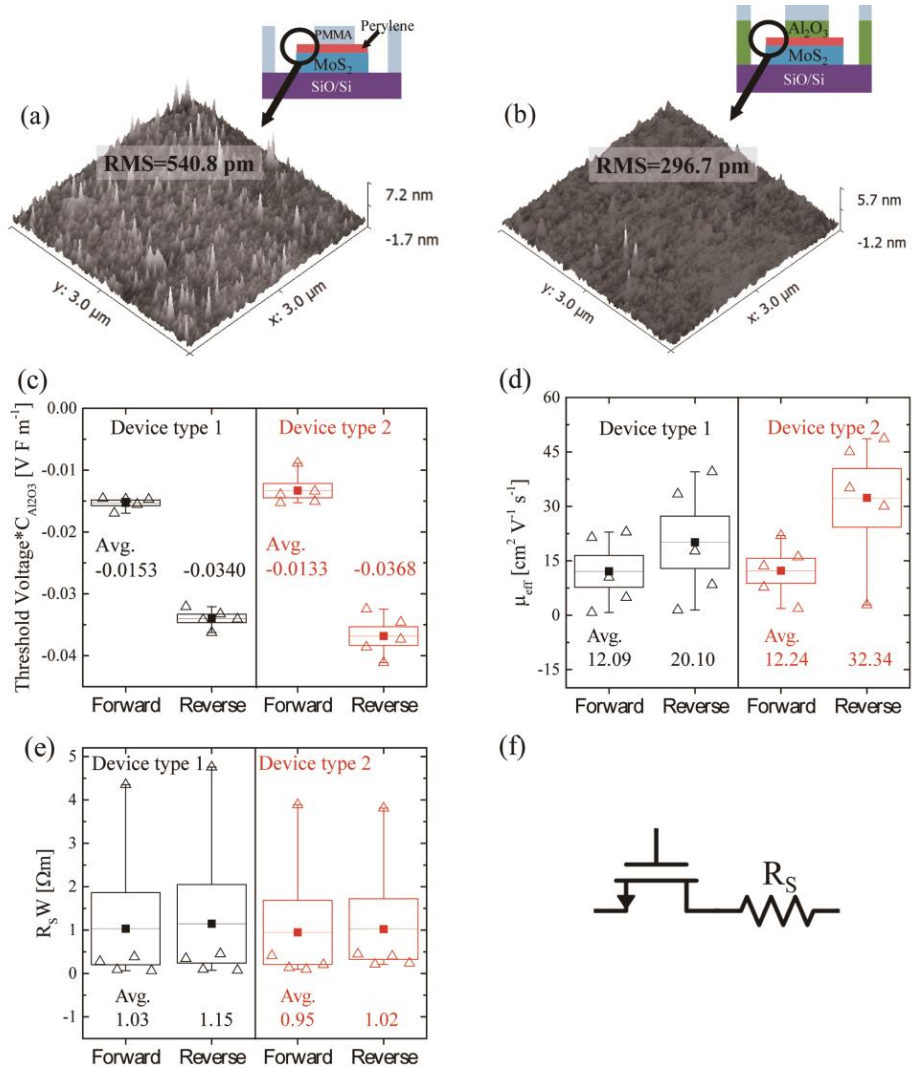

**Figure 7.** AFM images of the device type 1 (a) and device type 2 (b) prior to the deposition of the source-drain electrode. (c) Threshold voltage multiplied by $C_{Al2O3}$, (d) field-effect mobilities and (e) contact resistance multiplied by the channel width. Distributions of five samples of device type 1 (black) and device type 2 (red) for the $V_{TG}$ sweep directions. Open triangles denote the data for each sample. Filled squares denote the average of five samples. Boxes correspond to the standard error of samples. (f) Equivalent circuit with a serial resistor connected to the transistor. Contact resistance is extracted by using this model.

The main difference between device type 1 and 2 is that in the latter case the functionalized MoS$_2$ channel was not in contact with the resist during the fabrication. It is well known that resist residues can remain on the film surface after development and lift-off processes. To investigate this, the MoS$_2$ surface was characterized by AFM during fabrication. Figure 7(a) shows the functionalized MoS$_2$ surface after the development of the polymer resists during the fabrication of device type 1 (step 2 in Figure 3(a)). A relatively rough surface with a root-mean-square (RMS) roughness of 0.54 nm was observed. This roughness is likely due to polymer residues which remain in the channel area. In contrast, the functionalized MoS$_2$ surface after wet-etching of the contact area (step 3 in Figure 5(a)) of the device type 2 appears to be relatively flat (RMS roughness = 0.30 nm). Both the RMS and average values of the surface roughness of device type 1 clearly exceed those of device type 2. Thus one can deduce that while type 1 devices have polymer residues on the surface of the channel, type 2 devices have a relatively clean interface to the top gate dielectric. Thus we can tentatively attribute the second $g_m$ peak to the polymer residue in the channel area in device type 1, however additional experiments are required to analyze the exact mechanisms that cause the second $g_m$ peak.

To investigate the influence of resist residues on the device performance, five samples were fabricated and compared for each device type. Threshold voltage, mobilities and contact resistances of the devices are summarized in Figure 7(c), (d) and (e). Device type 1 (34 nm) and 2 (40 nm) has different Al$_2$O$_3$ thickness. Even applying same gate voltage, gate electric field is different depending on dielectric thickness. As considered to gate electric field, threshold voltage was multiplied by $C_{Al2O3}$. As shown in Figure 7(c), the hysteresis ($V_{TH,reverse}$-$V_{TH,forward}$) of device type 2 (0.0235 VFm$^{-1}$) increased by 26% compared to device type 1 (0.0187 VFm$^{-1}$). During the second deposition of Al$_2$O$_3$ of device type 2, interface states would form between the first and second ALD-Al$_2$O$_3$ layers and the overall quality of the Al$_2$O$_3$ of device type 2 would get worse, leading to large hysteresis compared to device type 1. As shown in Figure 7(d), the average $\mu_{FE}$ of device type 1 is lower than that of device type 2. In particular, the average $\mu_{FE}$ of the device type 2 (32.3 cm$^2$/Vs) for the reverse sweep increased by 61% compared to device type 1 (20.1 cm$^2$/Vs). This can be tentatively assigned to increased surface roughness and remote charge scattering. As shown in Figure 7(a), the functionalized MoS$_2$ surface of device type 1 has resist residues in the channel area, unlike device type 2. Even though the resist residues may not directly adhere to MoS$_2$ due to the perylene SAM, it can increase the surface roughness and act like a fixed charge inside the gate stack. This latter disturbance can cause scattering by remote surface scattering.[59,60] Further studies are needed to understand why the $\mu_{FE}$ of both device types is more pronounced in the reverse sweep than in the forward sweep in Figure 7(d). The effect of resist residues on the contact resistance of the device was also investigated. As shown in Figure 7(e), in the high $V_{TG}$ region in Figure 4(b) and Figure 6(b), the current is saturated due to the influence of contact resistance ($R_s$). $R_s$ is extracted by using an equivalent circuit model with a resistor serially connected to the transistor as shown in Figure 7(f). $R_s$ is obtained by using the graphical method[61] and the following equation:

$$I_{DS} = \mu_{FE} C_{ox} \frac{W}{L} \left[ (V_{TG} - V_{TH})(V_{DS} - I_{DS} R_s) - \frac{1}{2}(V_{DS} - I_{DS} R_s)^2 \right] \quad (2)$$

At $V_{DS}$=1 V, the $R_s$ along each sweep direction was calculated by using the transfer characteristics. Different channel widths (W) were considered for each device, and the width was multiplied by $R_s$. The distribution of the $R_s$W of device type 1 and 2 is shown in the Figure 7(e). The average $R_s$W of device type 2 (0.95 Ωm) for forward sweep reduced by 8% compared to device type 1 (1.03 Ωm), and the average $R_s$W of device type 2 (1.02 Ωm) for reverse sweep reduced by 11% compared to device type 1 (1.15 Ωm). This strongly suggests that the absence of polymer residues in the source and drain contact regions reduces the contact resistance, but the effect of polymer residues on mobility exceeds that of the contact resistance. In table 1 the best mobilities for various top-gated FETs with CVD-grown MoS$_2$ channels are listed. Compared to mobility values in literature, device type 2 of our work exhibits the comparable mobility. Thus the process flows with perylene bisimide functionalisation can be suggested as a viable route for MoS$_2$ integration, leading to reliable dielectric deposition with reproducible results and high mobilities.

Table 1. Summary of reported effective mobilities for the top-gated CVD MoS$_2$ FETs.

| Dielectric | Substrate | Atmosphere | $\mu_{FE}$ [cm$^2$/Vs] | Ref. |
|---|---|---|---|---|
| Al$_2$O$_3$ | SiO$_2$ | air | ~2 | [62] |
| AlN/Al$_2$O$_3$ | SiO$_2$ | air | 3.3 | [63] |
| HfO$_2$ | SiO$_2$ | air | 11 | [64] |
| Al$_2$O$_3$ | Si$_3$N$_4$ | ~10$^{-6}$ Torr | 24 | [65] |
| Al$_2$O$_3$ | SiO$_2$ | 3.7 Torr | ~33.4 | Device type 1 |
| Al$_2$O$_3$ | SiO$_2$ | 3.7 Torr | ~48.7 | Device type 2 |
| SiO$_2$/HfO$_2$ | SiO$_2$ | - | 42.3 | [66] |
| HfO$_x$ | SiO$_2$ | air | 55 | [67] |

**Conclusion**

In this study, top-gated FETs with CVD-grown MoS$_2$ were fabricated. A non-covalent perylene bisimide functionalization was used to facilitate ALD of Al$_2$O$_3$ as a dielectric and passivation layer. Perylene bisimide was simply deposited on MoS$_2$ by drop-casting at room temperature. Furthermore, we were able to define vertical contacts to the MoS$_2$ channels, yielding fully-encapsulated MoS$_2$ FETs. Perylene bisimide was non-destructively attached to MoS$_2$ and led to improved device performance as revealed by surface characterization and electrical measurements. The field-effect mobility for the MoS$_2$ FETs was found to be 48.7 cm$^2$/Vs. Thus our work suggests that non-covalent functionalisation is a viable strategy to fabricate devices with monolayer 2D materials. Additionally, we investigated the effect of resist residues on field-effect mobility and contact resistance. This study represents a significant step towards the fabrication of reproducible TMD-based devices, with interface engineering for passivation and dielectric deposition and contact formation as well as better understanding of the effects of polymer residues.

**Acknowledgements**

The authors wish to thank Dr. Sojeong Park and Dr. Christian Wirtz. This research is supported by National Research Foundation of Korea - Ministry of Science and ICT (NRF-2017M3A7B4049119) and by Multi-Ministry Collaborative R&D Program through the National Research Foundation of Korea funded by KNPA, MSIT, MOTIE, ME, NFA(NRF-2017M3D9A1073924). We further acknowledge support from the Science Foundation Ireland (SFI) for grants 15/SIRG/3329, 12/RC/2278 and 15/IA/3131, and the European Union Seventh Framework Program under Graphene Flagship, 604391.

**Supporting Information**

# Optimized single-layer MoS$_2$ field effect transistors by non-covalent functionalisation

HyunJeong Kim,‡[1,2] WungYeon Kim,‡[1,2] Maria O'Brien,[1] Niall McEvoy,[1] Chanyoung Yim,[3] Mario Marcia,[4] Frank Hauke,[4] Andreas Hirsch,[4] Gyu-Tae Kim,[2] and Georg S. Duesberg[1,3,a]

[1] CRANN&AMBER Centres and School of Chemistry, Trinity College Dublin, Dublin 2, Ireland

[2] School of Electrical Engineering, Korea University, Seoul 02841, South Korea

[3] Institute of Physics, EIT 2, Faculty of Electrical Engineering and Information Technology, Universität der Bundeswehr München, Neubiberg, Germany

[4] Chair of Organic Chemistry II, Friedrich-Alexander University of Erlangen-Nürnberg, Nikolaus Fiebiger Str. 10, 91058 Erlangen, Germany

E-mail: duesberg@unibw.de

‡These authors contributed equally to this work.

**Thicknesses of Al$_2$O$_3$ of top-gated FETs**

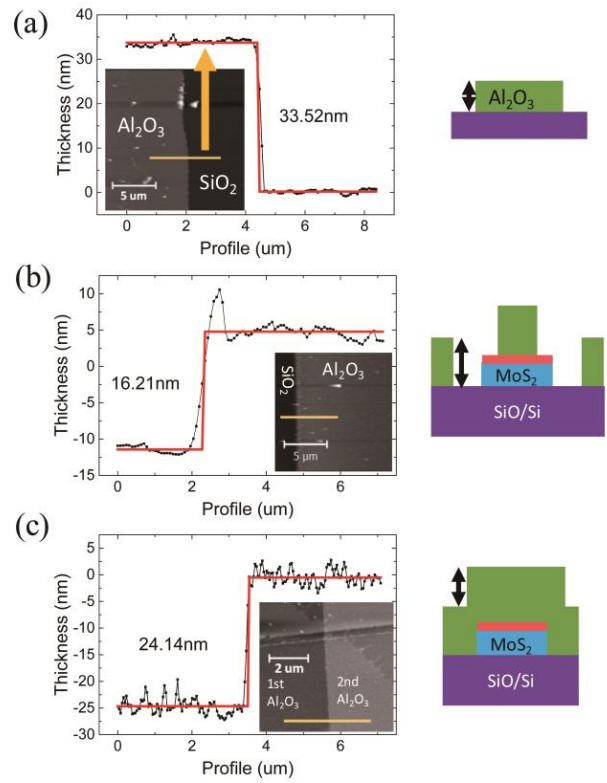

**Figure S1.** (a) AFM height profile of ALD-Al$_2$O$_3$ of device type 1. (b) AFM height profile after the first ALD-Al$_2$O$_3$ deposition, (After Al$_2$O$_3$ etch (Figure 5(a) step 3 and an additional removal of the PMMA) of device type 2. (c) AFM height profile after second ALD-Al$_2$O$_3$ deposition of device type 2.

**Back-gated CVD MoS$_2$ FETs**

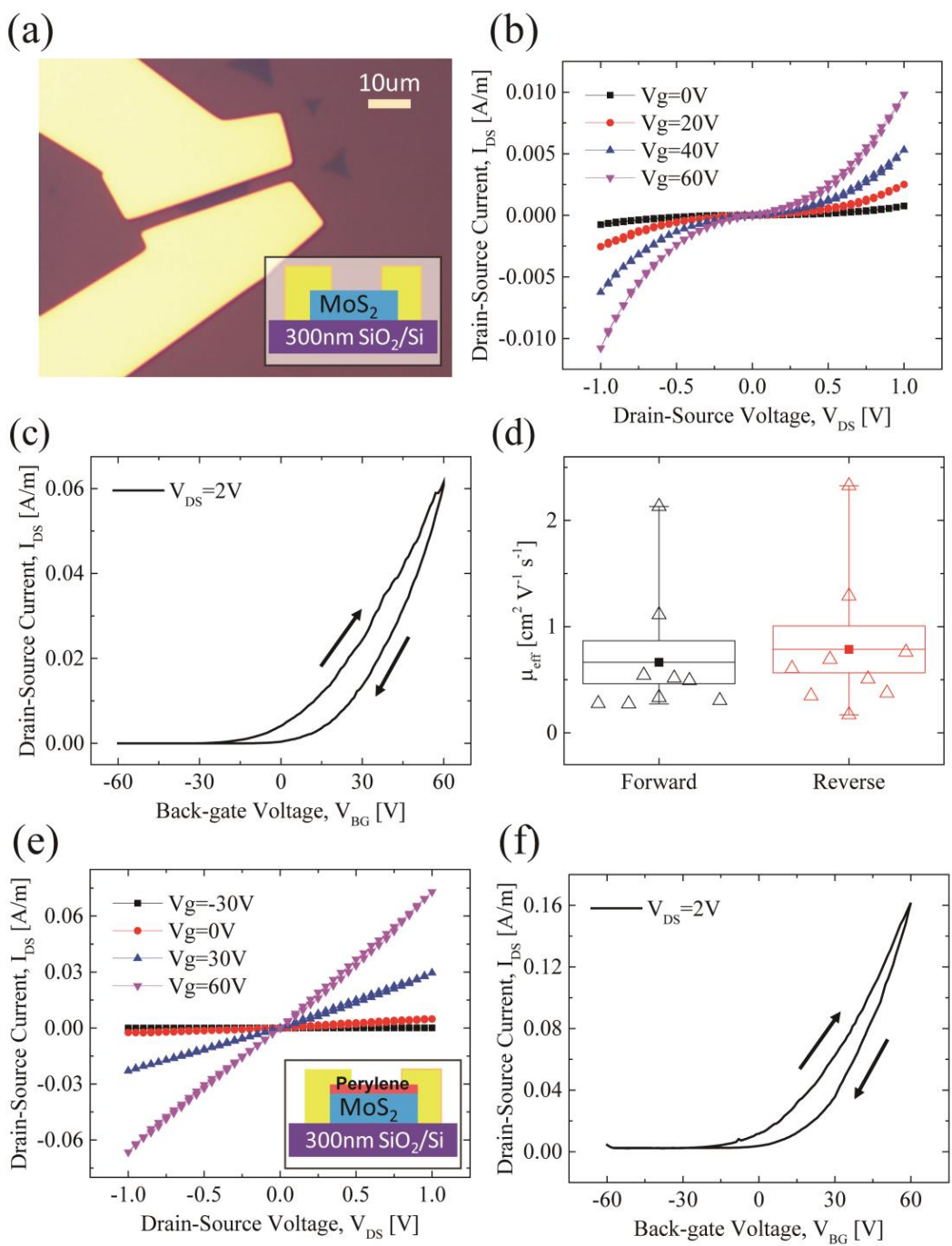

**Figure S2.** (a) Optical image of a back-gated CVD $MoS_2$ FET. The inset shows schematic of the device. (b) Output characteristics of the device shown in (a). (c) Transfer characteristics for the back-gated bias range of ±60 V at $V_{DS}$=2 V of the device. (d) Distribution of field-effect mobilities for forward sweep (black) and reverse sweep (red), this was extracted from nine devices. Open triangles denote the data for each sample. Filled squares denote the average of nine samples. Boxes correspond to the standard error of samples. (e) Output characteristics of a back-gated CVD $MoS_2$ FET with functionalisation. (f) Transfer characteristics of the device with functionalized $MoS_2$.

Figure S2(b) shows the $I_{DS}$-$V_{DS}$ output characteristics at various back-gate voltages ($V_{BG}$). Figure S2(c) shows the typical shapes of $I_{DS}$-$V_{GS}$ transfer characteristics with clockwise hysteresis. In figure S2(d), the average field-effect mobility of the devices is shown to be 0.664 cm$^2$/Vs for forward sweep and 0.786 cm$^2$/Vs for reverse sweep. Figure S2(e) and (f) show the electrical characteristics of back-gated CVD MoS$_2$ FET with functionalisation by perylene bisimide. Perylene bisimide was deposited before patterning of source/drain electrodes on MoS$_2$. Except for the functionalisation, all other fabrication processes are the same as the back-gated MoS$_2$ FET of figure S2(a) – (d). For the back-gated device with functionalisation, figure S2(f) shows a similar clockwise hysteresis like figure S2(c). From the electrical characteristics of the back-gated FETs it is reasonable to infer that perylene bisimide does not significantly affect the hysteresis between MoS$_2$ and electrodes.